\documentclass[twocolumn,superscriptaddress,showpacs,preprintnumbers,amsmath,amssymb]{revtex4}

\usepackage{graphicx,color}
\usepackage{dcolumn}
\usepackage{bm}

\begin{document}


\title{Spin-isospin resonances: A self-consistent covariant description}

\author{Haozhao Liang}
\affiliation{State Key Laboratory of Nuclear Physics {\rm\&} Technology,
    School of Physics, Peking University, Beijing 100871, China}
\affiliation{Institut de Physique Nucl\'eaire, IN2P3-CNRS and Universit\'e Paris-Sud,
    F-91406 Orsay Cedex, France}

\author{Nguyen Van Giai}
\affiliation{Institut de Physique Nucl\'eaire, IN2P3-CNRS and Universit\'e Paris-Sud,
    F-91406 Orsay Cedex, France}

\author{Jie Meng}
\affiliation{State Key Laboratory of Nuclear Physics {\rm\&} Technology,
    School of Physics, Peking University, Beijing 100871, China}
 \affiliation{Department of Physics, University of Stellenbosch, Stellenbosch, South Africa}

\date{\today}

\begin{abstract}
For the first time a fully self-consistent charge-exchange
relativistic RPA based on the relativistic Hartree-Fock (RHF)
approach is established. The self-consistency is verified by the
so-called isobaric analog state (IAS) check. The excitation
properties and the non-energy weighted sum rules of two important
charge-exchange excitation modes, the Gamow-Teller resonance (GTR)
and the spin-dipole resonance (SDR), are well reproduced in the
doubly magic nuclei $^{48}$Ca, $^{90}$Zr and $^{208}$Pb without
readjustment of the particle-hole residual interaction. The dominant
contribution of the exchange diagrams is demonstrated.
\end{abstract}

\maketitle


At present, spin-isospin resonances become one of the central topics
in nuclear physics and astrophysics. Basically, a systematic pattern
of the energy and collectivity of these resonances could provide
direct information on the spin and isospin properties of the
in-medium nuclear interaction, and the equation of state of
asymmetric nuclear matter. Furthermore, a basic and critical
quantity in nuclear structure, neutron skin thickness, can be
determined indirectly by the sum rule of spin-dipole resonances
(SDR) \cite{Krasznahorkay99,Yako06} or the excitation energy spacing
between isobaric analog states (IAS) and Gamow-Teller resonances
(GTR) \cite{Vretenar03}. More generally, spin-isospin resonances
allow one to attack other kinds of problems outside the realm of
nuclear structure, like the description of neutron star and
supernova evolutions, the $\beta$-decay of nuclei which lie on the
r-process path of stellar nucleosynthesis \cite{Engel99,Niksic05},
even the existence of exotic odd-odd nuclei \cite{Byelikov07} and
the efficiency of a solar neutrino detector \cite{Bhattacharya00}.


It was realized long ago that the Random Phase Approximation (RPA)
is an appropriate microscopic approach for charge-exchange giant
resonances \cite{Halbleib67,Engelbrecht70}. The importance of full
self-consistency was stressed \cite{Engelbrecht70}, and Skyrme-RPA
calculations of charge-exchange modes exist for about 30 years
\cite{Auerbach81}. Recently, a fully self-consistent charge-exchange
Skyrme-QRPA model has been developed \cite{Fracasso05}.
Self-consistency is an extremely important requirement for the
analysis of long isotopic chains extending towards the drip lines.
On the relativistic side, so far the charge-exchange (Q)RPA model
based on the relativistic mean field (RMF) theory has been developed
\cite{Conti98, Vretenar03, Ma04, Paar04}.

However, the self-consistency of the RMF+RPA is not completely
fulfilled for the following reasons. First, the isovector pion plays
an important role in the relativistic description of spin-isospin
resonances. Because of the parity conservation this degree of
freedom is absent in the ground-state description under the Hartree
approximation. Therefore, the pion is out of control in this
best-fitting effective field theory. Second, to cancel the contact
interaction coming from the pseudovector pion-nucleon coupling, a
zero-range counter-term is needed with the strength $g'=1/3$ exactly
\cite{Bouyssy87}. However, in order to reproduce the excitation
energies of the GTR, $g'$ must be treated as an adjustable parameter
in RMF+RPA model with the value $g'\approx 0.6$ \cite{Conti98,
Paar04}.


In this Letter, for the first time a fully self-consistent
charge-exchange relativistic RPA model is established, based on the
relativistic Hartree-Fock (RHF) approach \cite{Bouyssy87,Long06}.
The two major advantages of this RHF+RPA approach are that the pion
is included in both the ground-state description and the
particle-hole (p-h) residual interaction, and that the zero-range
pionic counter-term with $g'=1/3$ is maintained self-consistently.
Without any adjusted p-h residual interaction or re-fitting process,
we expect the present RHF+RPA approach to be reliable and to have
predictive power.

For a self-consistent calculation, the RPA p-h residual interaction
must be derived from the same Lagrangian as ground-state
\cite{Bouyssy87,Long06}. In the one-meson exchange picture, the
interactions are generated by $\sigma$-, $\omega$-, $\rho$-, and
$\pi$-meson exchanges. The nucleon-nucleon interactions read
\begin{subequations}\label{ph}
\begin{eqnarray}
    V_\sigma(1,2) &=& -[g_\sigma\gamma_0]_1[g_\sigma\gamma_0]_2D_\sigma(1,2),\label{ph sigma}\\
    V_\omega(1,2) &=& [g_\omega\gamma_0\gamma^\mu]_1[g_\omega\gamma_0\gamma_\mu]_2D_\omega(1,2),\label{ph omega}\\
    V_\rho(1,2) &=& [g_\rho\gamma_0\gamma^\mu\vec\tau]_1\cdot [g_\rho\gamma_0\gamma_\mu\vec\tau]_2D_\rho(1,2),\label{ph rho}\\
    V_\pi(1,2) &=& -[\frac{f_\pi}{m_\pi}\vec\tau\gamma_0\gamma_5\gamma^k\partial_k]_1\cdot
         [\frac{f_\pi}{m_\pi}\vec\tau\gamma_0\gamma_5\gamma^l\partial_l]_2\nonumber\\
         && \times D_\pi(1,2),\label{ph pion}
\end{eqnarray}
\end{subequations}
where $D_i(1,2)$ denotes the finite-range Yukawa type propagator
\begin{equation}\label{Yukawa}
    D_i(1,2) = \frac{1}{4\pi}\frac{e^{-m_i|\boldsymbol r_1-\boldsymbol r_2|}}
        {|\boldsymbol r_1-\boldsymbol r_2|}.
\end{equation}
As discussed before, a zero-range pionic counter-term with $g'=1/3$
must be included,
\begin{equation}\label{ph counter}
    V_{\pi\delta}(1,2)=g'[\frac{f_\pi}{m_\pi}\vec\tau\gamma_0\gamma_5\boldsymbol\gamma]_1\cdot
         [\frac{f_\pi}{m_\pi}\vec\tau\gamma_0\gamma_5\boldsymbol\gamma]_2
         \delta(\boldsymbol r_1-\boldsymbol r_2).
\end{equation}
In the present RHF+RPA framework, both direct and exchange terms
must be taken into account and therefore the isoscalar mesons also
play their role in spin-isospin resonances via the exchange terms.
This is another distinct difference from the RMF+RPA model.


In order to verify the model self-consistency, we perform a
so-called IAS check according to the following property about IAS
degeneracy: it is expected that the IAS would be degenerate with its
isobaric multiplet partners if the nuclear Hamiltonian commutes with
the isospin lowering $T_-$ and raising $T_+$ operators, which is
true when the Coulomb field is switched off. While this degeneracy
is broken by the mean field approximation, it can be restored by the
self-consistent RPA calculation \cite{Engelbrecht70}. As an example
we calculate the IAS in $^{208}$Pb with the parametrization PKO1
\cite{Long06} and we find the unperturbed excitations between
$-10.46$ MeV and $-8.96$ MeV when the Coulomb interaction is put to
zero, thus showing the isospin symmetry breaking. Then, the RHF+RPA
calculation leads to $E_{\mbox{\scriptsize IAS}}=4$ keV within the
single-particle energy truncation $[-M, M+80\ \mbox{MeV}]$. This
restoration of the IAS degeneracy indicates that the present
approach is fully self-consistent. Furthermore, it should be
emphasized that the pion also plays its role in this restoration
process. Therefore, the coefficient $g'$ is not a free parameter.
Changing the value of $g'$ would destroy the symmetry restoration
process. For example, $g'=0$ leads to $E_{\mbox{\scriptsize
IAS}}=-801$ keV.

\begin{table*}
\caption{GTR excitation energies in MeV and strength in percentage
        of the $3(N-Z)$ sum rule within the RHF+RPA framework.
        Experimental \cite{Anderson85,Bainum80,Wakasa97,Horen80,Akimune95}
        and RMF+RPA \cite{Paar04} results are given for comparison.
        \label{Tab:GTR}}
\begin{ruledtabular}
    \begin{tabular}{cccccccc}
&&\multicolumn{2}{c}{$^{48}$Ca} &\multicolumn{2}{c}{$^{90}$Zr}
            &\multicolumn{2}{c}{$^{208}$Pb} \\
        && energy & strength & energy & strength & energy & strength\\
        \hline
        experiment && $\sim10.5$&$35$ &$15.6\pm0.3$&$28$&$19.2\pm0.2$&$60\mbox{-}70$\\
        \hline
        RHF+RPA & PKO1&10.72&69.4&15.80&68.1&18.15&65.6\\
        &PKO2&10.83&66.7&15.99&66.3&18.20&60.5\\
        &PKO3&10.42&70.7&15.71&68.9&18.14&67.7\\ \hline
        RMF+RPA & DD-ME1&10.28&72.5&15.81&71.0&19.19&70.6\\
    \end{tabular}
\end{ruledtabular}
\end{table*}

\begin{figure}
\includegraphics[width=0.45\textwidth]{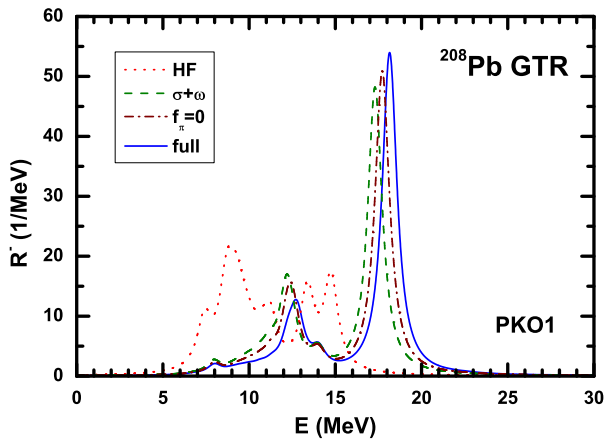}
\caption{(color online) Strength distribution of GTR in $^{208}$Pb calculated by
RHF+RPA with PKO1 (solid line).
    The unperturbed (HF) strength (dotted line),
    the calculation with only $\sigma+\omega$ p-h residual interaction (dashed line),
    and the calculation excluding pion ($f_\pi=0$)
    in the p-h residual interaction (dash-dotted line)
    are also shown. A Lorentzian smearing parameter $\Gamma=1$ MeV
    is used.
    \label{Fig:PbGTR}}
\end{figure}

Taking the doubly magic nuclei $^{48}$Ca, $^{90}$Zr and $^{208}$Pb
as test cases, and using the Gamow-Teller (GT) operator $F_\pm^{\rm
GT}=\sum_i\boldsymbol\sigma(i)\tau_\pm(i)$, the excitation energies
and strengths calculated with the fully self-consistent RHF+RPA
approach using the parametrizations PKO1, PKO2, PKO3
\cite{Long06set} are summarized in Table \ref{Tab:GTR}. These three
parametrizations correspond to different sets of coupling strengths
and meson masses in Eqs.~(\ref{ph})-(\ref{ph counter}). In PKO2 the
pion is not included (i.e. $f_\pi=0$), whereas PKO1 and PKO3 have
different constraints on the density dependence of $f_\pi$. A good
agreement with empirical energies is obtained without any
re-adjusted parameter. All calculated strengths correspond to the
main peak and contain 60-70\% of the Ikeda sum rule.

We can understand the different physical mechanisms between the
present RHF+RPA and other RMF+RPA approaches by the following
analysis. On the one hand, it has been shown that the $\pi NN$
interaction and its zero-range counter-term ($g'\approx0.6$) are the
dominant ingredients in p-h residual interaction for the GT mode in
the case of RMF+RPA \cite{Conti98,Paar04}. On the other hand, in the
present RHF+RPA calculations, three parametrizations PKO1, PKO2 and
PKO3 lead to similar results for the GTR excitation energies. It
should be emphasized that the pion is not included in PKO2. This
hints to the fact that the pion interaction is not the only dominant
ingredient for the GT excitations in this framework. The GT strength
distribution in $^{208}$Pb with PKO1 is shown in
Fig.~\ref{Fig:PbGTR}. It is compared with the calculation in which
the pion is excluded ($f_\pi=0$) in the p-h channel, the calculation
including only $\sigma+\omega$ p-h residual interactions, and the
unperturbed (HF) case. One can conclude that the isoscalar $\sigma$-
and $\omega$-mesons play an essential role via the exchange terms,
whereas the pion just stands on a marginal position in determining
the GTR strength distribution.


\begin{table*}
\caption{Ikeda sum rule values from Fermi ($S^{\rm{F}}$) and Dirac
    ($S^{\rm{D}}$) sectors.
    $S_-$, $S_+$ are the sum rule values of the $T_-$ and $T_+$ channels, respectively.
    The reduction factor, $1-(S^{\rm{F}}_- - S^{\rm{F}}_+)/(S_- - S_+)$,
    is given in the last column.
    \label{Tab:GTRReduction}}
\begin{ruledtabular}
    \begin{tabular}{ccccccccc}
        &&$S_-^{\rm{F}}$&$S_-^{\rm{D}}$&$S_+^{\rm{F}}$&$S_+^{\rm{D}}$
        & $S_-^{\rm{F}}-S_+^{\rm{F}}$& $S_--S_+$ & reduction\\ \hline
        $^{48}$Ca
        &PKO1 &22.67 &4.23 &0.10 &2.83 &22.57 &23.97 &5.9\%\\
        &PKO2 &22.67 &4.22 &0.15 &2.77 &22.53 &23.98 &6.1\%\\
        &PKO3 &22.66 &4.24 &0.13 &2.80 &22.53 &23.97 &6.0\%\\
        \hline
        $^{90}$Zr
        &PKO1 &28.22 &8.08 &0.32 &5.99 &27.91 &29.99 &7.0\%\\
        &PKO2 &28.29 &7.98 &0.41 &5.88 &27.87 &29.97 &7.0\%\\
        &PKO3 &28.23 &8.07 &0.36 &5.97 &27.86 &29.96 &7.0\%\\
        \hline
        $^{208}$Pb
        &PKO1 &122.94 &21.54 &0.51 &11.98 &122.43 &131.99 &7.2\%\\
        &PKO2 &123.05 &21.50 &0.83 &11.74 &122.22 &131.98 &7.4\%\\
        &PKO3 &122.77 &21.87 &0.66 &12.00 &122.11 &131.99 &7.5\%\\
    \end{tabular}
\end{ruledtabular}
\end{table*}

The relativistic RPA is equivalent to the time-dependent RMF in the
small amplitude limit if the p-h configuration space includes not
only the pairs formed from the occupied and unoccupied Fermi states
but also the pairs formed from the Dirac states and occupied Fermi
states \cite{Ring01}. Based on this idea, a relativistic reduction
mechanism of the Gamow-Teller strength due to the effects of the
Dirac sea states was pointed out \cite{Kurasawa03}. This kind of
reduction mechanism appears in both nuclear matter \cite{Kurasawa03}
and finite nuclei \cite{Ma04,Paar04}. In Table
\ref{Tab:GTRReduction}, we show the contributions to the Ikeda sum
rule values coming from the Fermi ($S^{\rm{F}}$) and Dirac
($S^{\rm{D}}$) sectors of both $T_\pm$ channels. It is explicitly
shown that the Ikeda sum rule
\begin{equation}\label{Ikeda sum rule}
    S^{\rm GT}_--S^{\rm GT}_+=3(N-Z)
\end{equation}
can be 100\% exhausted only when the effects of the Dirac sea are
included. The reduction factors, $1-(S^{\rm{F}}_- -
S^{\rm{F}}_+)/(S_- - S_+)$, of $^{48}$Ca, $^{90}$Zr, $^{208}$Pb by
the present self-consistent approach indicate to which extent the
antinucleon degrees of freedom play a role.


\begin{figure}
\includegraphics[width=0.45\textwidth]{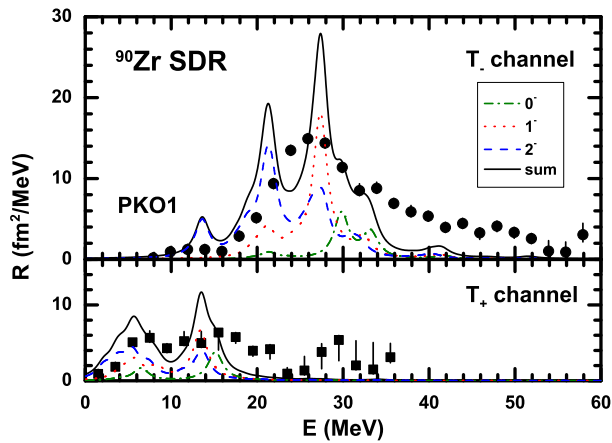}
\caption{(color online) Strength distributions in both $T_\pm$ channels of the SDR
    in $^{90}$Zr calculated by RHF+RPA with PKO1.
    The dash-dotted, dotted, dashed lines show the 0-, 1-, 2-
    contributions respectively,
    while the solid line shows their sum.
    A Lorentzian smearing parameter $\Gamma=2$ MeV is used.
    The experimental data are shown as filled symbols \cite{Yako06}.
    \label{Fig:ZrSDR}}
\end{figure}

\begin{table*}
\caption{SD sum rule values and neutron skin thickness of $^{90}$Zr, $^{208}$Pb
    in RHF+RPA approach.
    Neutron and proton rms radii and corresponding data
    from SD experiment \cite{Yako06} are given for comparison.
    $S(r_n,r_p)$ stands for the RHS of Eq. (\ref{SD sum rule}).
    \label{Tab:SDR}}
\begin{ruledtabular}
    \begin{tabular}{ccccccccc}
        && $r_n$ (fm) & $r_p$ (fm) & $\delta_{np}$ (fm)& $S(r_n,r_p)$ (fm$^2$)
        & $S_-^{\rm{F}}-S_+^{\rm{F}}$ (fm$^2$)& $S_--S_+$ (fm$^2$)& reduction\\ \hline
        $^{90}$Zr
        &SD exp. &$4.26\pm0.04$ &$4.19\pm0.01$ &$0.07\pm0.04$ & & &$148\pm12$\\
        \cline{2-9}
        &PKO1 &4.280 &4.188 &0.092 &153.5 &143.8 &153.6 &6.4\% \\
        &PKO2 &4.264 &4.184 &0.080 &149.6 &139.7 &149.4 &6.5\% \\
        &PKO3 &4.271 &4.192 &0.079 &149.8 &140.3 &149.9 &6.5\% \\ \hline
        $^{208}$Pb
        &PKO1 &5.691 &5.457 &0.234 &1174 &1111 &1174 &5.4\% \\
        &PKO2 &5.655 &5.461 &0.194 &1134 &1071 &1135 &5.6\% \\
        &PKO3 &5.658 &5.456 &0.202 &1141 &1077 &1141 &5.6\% \\
    \end{tabular}
\end{ruledtabular}
\end{table*}

It has been proposed that the neutron skin thickness could be
extracted via the spin-dipole sum rule\cite{Krasznahorkay99},
\begin{equation}\label{SD sum rule}
    S^{\rm SD}_- - S^{\rm SD}_+ = \frac{9}{4\pi}(N\left< r^2\right>_n - Z\left< r^2\right>_p),
\end{equation}
with the SD operator $F_\pm^{\rm SD}=\sum_i
[r_iY_1(i)\otimes\boldsymbol\sigma(i)]_{J=0,1,2}\tau_\pm(i)$. While
experimental results in both $^{90}$Zr$(p,n)$ \cite{Wakasa97} and
$^{90}$Zr$(n,p)$ \cite{Yako05} channels are now available, the SDR
has become a hot topic.

Both $T_-$ and $T_+$ channels of the SD strength distributions are
shown in Fig.~\ref{Fig:ZrSDR}. Here, the horizontal axis is the
excitation energy measured from the ground-state of the parent
nucleus $^{90}$Zr. The RHF+RPA calculations reproduce the strength
distributions up to the giant resonance region in both channels
without any quenching factor, even though they show a more
pronounced structure than the experimental spectra. Especially the
dominant resonance structure centered at $E_x\approx27$ MeV in the
$T_-$ channel is well reproduced. It is also found that the three
components of the SDR follow the same energy hierarchy
$E(2^-)<E(1^-)<E(0^-)$ as that found in the recent self-consistent
Skyrme-RPA calculations \cite{Sagawa07,Fracasso07}. Since the 1p-1h
configuration space of the RPA calculations is restricted, the
discrepancy beyond the giant resonance region can be understood.
Furthermore, in the first part of Table \ref{Tab:SDR} the neutron
and proton rms radii, the neutron skin thickness and the SD sum rule
value of $^{90}$Zr are listed. They all agree with the data within
the experimental accuracy.

For the spin-dipole sum rule values, it is worth to note that a
reduction due to the effects of the antinucleon degree of freedom in
analogy with the GTR case is found. The present calculations show
that this reduction is around 6.5\% in $^{90}$Zr.

\begin{figure}
\includegraphics[width=0.40\textwidth]{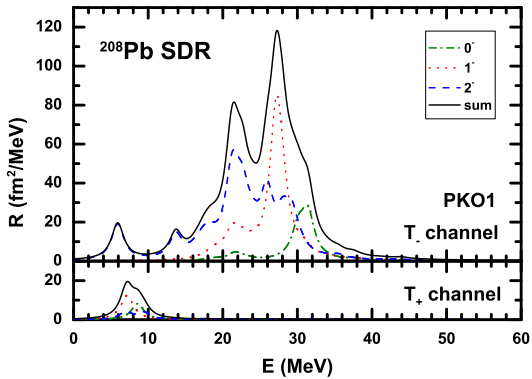}
\caption{(color online) Same as Fig.~\ref{Fig:ZrSDR}, but for the nucleus
$^{208}$Pb. \label{Fig:PbSDR}}
\end{figure}

Since the neutron skin thickness of $^{208}$Pb is important in many
aspects of nuclear physics and astrophysics, it is worthwhile to
investigate the SDR properties of $^{208}$Pb. From
Fig.~\ref{Fig:PbSDR} and the second part of Table \ref{Tab:SDR}, the
present approach predicts a dominant resonance structure at $E =
20\sim30$ MeV in the $T_-$ channel and a small bump at $E = 5\sim10$
MeV in the $T_+$ channel. Furthermore, a reduction factor of about
5.5\% is predicted by the three different parametrizations.


In conclusion, for the first time, a fully self-consistent
charge-exchange relativistic RPA model based on the RHF approach is
established. The IAS degeneracy broken by the RHF approximation can
be accurately restored in the present self-consistent RPA
calculations. Compared with RMF+RPA, the isoscalar mesons ($\sigma$,
$\omega$) are found to play an essential role in spin-isospin
resonances via the exchange terms. The GTR excitation energies and
their strengths can be reproduced in the present self-consistent RPA
calculation while maintaining $g'=1/3$ in the contact counter-term.
The SDR strength distributions up to the giant resonance region in
both channels are well reproduced in $^{90}$Zr. Furthermore, a
spin-dipole sum rule reduction mechanism due to the effects of the
Dirac sea is found. The SD reduction factor for $^{90}$Zr is around
$6.5\%$. Finally, the dominant structures of the SD strength
distribution in $^{208}$Pb are shown and a strength reduction of
about 5.5\% is obtained.


This work is partly supported by Major State Basic Research
Developing Program 2007CB815000, the National Natural Science
Foundation of China under Grant Nos. 10435010, 10775004 and
10221003, the European Community project Asia-Europe Link in Nuclear
Physics and Astrophysics CN/Asia-Link 008 (94791), and the
CNRS(France) - NSFC(China) PICS program no. 3473.


\begin{thebibliography}{99}
    \bibitem{Krasznahorkay99} A. Krasznahorkay et al., Phys. Rev. Lett. {\bf 82}, 3216 (1999).
    \bibitem{Yako06} K. Yako, H. Sagawa, H. Sakai, Phys. Rev. C {\bf 74}, 051303(R) (2006).
    \bibitem{Vretenar03} D. Vretenar, N. Paar, T. Nik\v{s}i\'{c}, P. Ring,
        Phys. Rev. Lett. {\bf 91}, 262502 (2003).
    \bibitem{Engel99} J. Engel, M. Bender, J. Dobaczewski, W. Nazarewicz, R. Surman,
        Phys. Rev. C {\bf 60}, 014302 (1999).
    \bibitem{Niksic05} T. Nik\v{s}i\'{c}, T. Marketin, D. Vretenar, N. Paar, P. Ring,
        Phys. Rev. C {\bf 71}, 014308 (2005).
    \bibitem{Byelikov07} A. Byelikov et al., Phys. Rev. Lett. {\bf 98}, 082501 (2007).
    \bibitem{Bhattacharya00} M. Bhattacharya et al., Phys. Rev. Lett. {\bf 85}, 4446 (2000).
    \bibitem{Halbleib67} J. A. Halbleib, R. A. Sorensen, Nucl. Phys. A {\bf 98}, 542 (1967).
    \bibitem{Engelbrecht70} C.A. Engelbrecht and R. H. Lemmer,
        Phys. Rev. Lett. {\bf 24}, 607 (1970).
    \bibitem{Auerbach81} N. Auerbach, A. Klein and N. Van Giai,
        Phys. Lett. B {\bf 106}, 347 (1981).
    \bibitem{Fracasso05} S. Fracasso, G. Col\`{o}, Phys. Rev. C {\bf 72}, 064310 (2005).
    \bibitem{Conti98} C. De Conti, A.P. Gale\~{a}o, F. Krmpoti\'{c},
        Phys. Lett. B {\bf 444}, 14 (1998).
    \bibitem{Ma04} Z.Y. Ma, B.Q. Chen, N. Van Giai, T. Suzuki,
        Eur. Phys. J. A {\bf 20}, 429 (2004).
    \bibitem{Paar04} N. Paar, T. Nik\v{s}i\'{c}, D. Vretenar, P. Ring,
        Phys. Rev. C {\bf 69}, 054303 (2004).
    \bibitem{Bouyssy87} A. Bouyssy, J.F. Mathiot, N. Van Giai, S. Marcos,
        Phys. Rev. C {\bf 36}, 380 (1987).
    \bibitem{Long06} W.H. Long, N. Van Giai, J. Meng, Phys. Lett. B {\bf 640}, 150 (2006).
    \bibitem{Long06set} W.H. Long, N. Van Giai and J. Meng, arXiv:nucl-th/0608009.
    \bibitem{Ring01} P. Ring, Z.Y. Ma, N. Van Giai, D. Vretenar, A. Wandelt, L.G. Cao,
        Nucl. Phys. A {\bf 694}, 249 (2001).
    \bibitem{Kurasawa03} H. Kurasawa, T. Suzuki, N. Van Giai,
        Phys. Rev. Lett. {\bf 91}, 062501 (2003); Phys. Rev. C {\bf 68}, 064311 (2003).
    \bibitem{Anderson85} B.D. Anderson et al., Phys. Rev. C {\bf 31}, 1161 (1985).
    \bibitem{Bainum80} D.E. Bainum et al., Phys. Rev. Lett. {\bf 44}, 1751 (1980).
    \bibitem{Wakasa97} T. Wakasa et al., Phys. Rev. C {\bf 55}, 2909 (1997).
    \bibitem{Horen80} D.J. Horen et al., Phys. Lett. B {\bf 95}, 27 (1980).
    \bibitem{Akimune95} H. Akimune et al., Phys. Rev. C {\bf 52}, 604 (1995).
    \bibitem{Yako05} K. Yako et al., Phys. Lett. B {\bf 615}, 193 (2005).
    \bibitem{Sagawa07} H. Sagawa, S. Yoshida, X. R. Zhou, K. Yako, H. Sakai,
        Phys. Rev. C {\bf 76}, 024301 (2007).
    \bibitem{Fracasso07} S. Fracasso, G. Col\`{o}, Phys. Rev. C {\bf 76}, 044307 (2007).
\end{thebibliography}
\end{document}